\documentclass[5p,english]{elsarticle}
\usepackage[OT1]{fontenc}
\usepackage[latin9]{inputenc}
\usepackage{amsmath}
\usepackage{amstext}
\usepackage{graphicx}

\makeatletter
\journal{Solid State Communications}
\IfFileExists{lmodern.sty}{\usepackage{lmodern}}{}

\usepackage{babel}
\biboptions{sort&compress}

\makeatother

\usepackage{babel}
\begin{document}

\title{Bell inequality and nonlocality in a two-dimensional mixed spin systems}

\author[rvt]{Z.Y.~Sun}

\ead{sunzhaoyu2012@gmail.com}

\author[rvt]{Y.Y.~Wu}

\ead{wuyuying0913@gmail.com}

\author[rvt]{H.L.~Huang\corref{cor1}}

\ead{hailin\_huang@whpu.edu.cn}

\author[BJ]{B.~Wang}

\ead{bowangphysics@gmail.com}

\cortext[cor1]{Corresponding author}

\address[rvt]{School of Electrical and Electronic Engineering, Wuhan Polytechnic
University, Wuhan 430000, China}

\address[BJ]{Department of Physics, Beijing Normal University, Beijing 100875,
China }

\address{}
\begin{abstract}
In this paper, we use Bell inequality and nonlocality to study the
bipartite correlation in an exactly soluble two-dimensional mixed
spin system. Bell inequality turns out to be a valuable detector for
phase transitions in this model. It can detect not only the quantum
phase transition, but also the thermal phase transitions, of the system.
The property of bipartite correlation in the system is also analyzed.
In the quantum anti-ferromagnetic phase, the Bell inequality is violated
thus nonlocality is present. It is interesting that the nonlocality
is enhanced by thermal fluctuation, and similar results have not been
observed in anti-ferromagnetic phase. In the ferromagnetic phase,
the quantum correlation turns out to be very novel, which cannot be
captured by entanglement or nonlocality.\end{abstract}
\maketitle

\section{Introduction}

Recently, quantum correlation in low-dimensional systems has attracted
much attention. One the one hand, quantum correlation plays a central
role in the formation of various condense matter phases at low temperatures.
It helps us to understand these phases and the so-called quantum phase
transitions(QPTs).\citep{BOOK} For example, for various models it
has been found quantum correlation is maximum or singular in the vicinity
of the transition points.\citep{Bell_inequalitiesQPTs_XXZ_model,Classical_correlation_and_quantum_discord_in_critical_systems,Correlation_nonlocalityQPTS_several_systems,QE4,Quantum_correlations_topological_QPT,Quantum_topological_QPT_at_microscopic_level}
On the other hand, quantum correlation is regarded as valuable resource
for quantum information communication and quantum computation.\citep{quantum computing,Quantum_Discord_A_Measure_of_the_Quantumness_of_Correlations,Unified_View_of_Quantum_and_Classical_Correlations}

The most famous features of quantum correlation are quantum entanglement
and nonlocality. Entanglement is defined on the separability-entanglement
paradigm.\citep{Unified_View_of_Quantum_and_Classical_Correlations,def_CC}
For example, the two-qubit pure state $|\uparrow_{1}\uparrow_{2}\rangle$
is separable, since it can be expressed as the product of the single-site
states $|\uparrow_{1}\uparrow_{2}\rangle=|\uparrow_{1}\rangle\otimes|\uparrow_{2}\rangle$.
However, most states cannot be expressed as such a product form, e.g.,
the superposition of two separable states $|\uparrow_{1}\uparrow_{2}\rangle+|\downarrow_{1}\downarrow_{2}\rangle$.
They are called entangled states. The entanglement measurement is
usually defined in the bipartite setting, such as entanglement concurrence
and entanglement entropy. Recently, it has been recognized that entanglement
can be generalized to multipartite settings.\citep{Hierarchies-of-Multipartite-Entanglement,multipartite-pure-states}

Besides entanglement, nonlocality is also an important feature of
quantum correlation, which is described by the violation of Bell inequalities\citep{Gisin_Entanglement_is_Nonlocal_twoQubitPureState,Bell_inequalities_in_Heisenberg_spin_chains,Bell_inequalitiesQPTs_XXZ_model,Bell_Bell_Inequalityies,Werner_Bell_Ineaulity_mixed_state}.
We consider some expression $\mathcal{B}(\hat{\rho})$, namely, a
function for the state $\hat{\rho}$. We will called it a Bell function.
For all the states described by a realistic local theory, we can always
identify the upper bound $\mathcal{B}_{0}$ for $\mathcal{B}(\hat{\rho})$.
In other words, for any local state $\hat{\rho}$, it should hold
\[
\mathcal{B}(\hat{\rho})\le\mathcal{B}_{0}.
\]
It is just a Bell inequality. For some state $\hat{\rho}$, suppose
$\mathcal{B}(\hat{\rho})$ turns out to be larger than $\mathcal{B}_{0}$,
we will say that the Bell inequality is violated, thus $\hat{\rho}$
cannot be characterized by any realistic local theory, i.e., $\hat{\rho}$
is a non-local state. A widely used Bell inequality is proposed by
Clauser, Horne, Shimony and Holt, namely, the CHSH inequality.\citep{CHSH}
Theoretically, numerical optimization can be used to identify the
upper bound $\mathcal{B}_{0}$. Fortunately, a closed analytical formula
of the Bell inequality for two-qubit states has been found by Horodecki.\citep{Horodecki_BFV_twoQubitState}

Nonlocality and entanglement were regarded as similar concepts for
years. However, an entangled state may not violate any Bell inequality.\citep{Nonlocality_entanglement_XY_model,Nonlocality_entanglement__qubit_systems}
For example, for several matrix product states, we found that the
two-qubit states do not violate Bell inequality but they are entangled.\citep{Bell_MPS_QPT}
Thus, nonlocality and entanglement capture different aspects of quantum
correlation. In fact, there even exists some kind of quantum correlation,
which cannot be characterized by nonlocality and entanglement, as
we will show in this paper. Quite recently, it has been found that
Bell function can be used to indicate QPTs for many one-dimensional
quantum systems.\citep{Correlation_nonlocalityQPTS_several_systems}
It is interesting that it even can be used for topological QPT\citep{Quantum_topological_QPT_at_microscopic_level,BFV_Topological_QPT}
and Kosterlitz-Thouless QPT\citep{Bell_inequalitiesQPTs_XXZ_model},
which are difficult to capture by traditional order parameters.

Previous studies about Bell inequality and nonlocality are limited
to QPTs in one-dimensional systems. Surprisingly, for various systems
it turns out that the Bell inequality is not violated.\citep{Bell_inequalitiesQPTs_XXZ_model,Correlation_nonlocalityQPTS_several_systems}
It is recently realized that such an unexpected result is related to
the monogamy trade-off obeyed by bipartite Bell function. As a
result, in translation invariant systems, nonlocality should not be
violated in general.\citep{Nonviolation-Bell-translation-invariant-systems}
Based on this consideration, two-dimensional systems would be more
suitable for preparing and then studying nonlocality. Firstly, two-dimensional
systems have complex topology and a perfect translation-invariance
symmetry is usually absent. Thus, the nonlocality can be present in
the systems. Secondly, two-dimensional systems undergo not only QPTs
at zero temperature, but also thermal phase transitions at finite
temperatures. The study of nonlocality in thermal phase transitions
would of course increase our understanding of quantum correlation
and phase transitions.

In this paper, we will study Bell inequality and nonlocality in a
two-dimensional Heisenberg-Ising mixed spin system.\citep{2D-mixed_model}
Firstly, it is exactly soluble, thus we can concentrate on physics
rather than mathematical calculation. Secondly, the model has a rich
phase diagram. At zero temperature the system has an ferromagnetic
(FM) phase and a quantum anti-ferromagnetic (QAF) phase, and a first-order
QPT happens between the two phases. At finite temperatures, ordered
ground states will be destroyed by thermal fluctuation, and the system
will enter the paramagnetic (PM) phase. As a result, the system will
undergo second-order thermal phase transition from the ordered magnetic
phases to PM phase at finite temperatures. The phase diagram of the
model is firstly identified by investigating the spin-spin correlation
functions.\citep{2D-mixed_model} Recently, the system is investigated
with the help of modern quantum information tools,\citep{SunEpl,CPL-fidelity}
and some issues remain unresolved. In the FM phase quantum entanglement
turns out to be absent, however, quantum correlation is found to be
present, indicated by the discord (discord measures all the quantum
correlation in a quantum state).\citep{SunEpl,Quantum_Discord_A_Measure_of_the_Quantumness_of_Correlations}
It is natural to ask, what is the intuitive form of the quantum correlation
in the FM phase of the model? Can it be characterized by nonlocality?
This is another motivation for us to investigate the nonlocality in
this model.

This paper is organized as follows. In Sec. 2, we briefly introduce
the model and basic formula. In Sec. 3, we study the relationship
between Bell function and various phase transitions in the model.
In Sec. 4, we discuss the properties of bipartite correlation in the
system by analyzing nonlocality and other measures such as entanglement.
A summary is given in Sec. 5.

\section{Basic formula}

\subsection{Horodecki's criterion}

Firstly, let's introduce Horodecki's criterion \citep{Horodecki_BFV_twoQubitState}
for nonlocality, which provides a closed analytic expression for the
Bell inequality for any two-qubit state $\hat{\rho}$. We defines
a matrix $\hat{\mathcal{L}}$ as
\begin{equation}
\mathcal{L}_{ij}(\hat{\rho})=\textrm{Tr}[\hat{\rho}\cdot\hat{\sigma}_{i}\otimes\hat{\sigma}_{j}],
\end{equation}
where $\hat{\sigma}_{1},\hat{\sigma}_{2},\hat{\sigma}_{3}$ are just
Pauli matrices. Then one finds the two largest eigenvalues of $\hat{\mathcal{L}}^{T}\hat{\mathcal{L}}$,
denoted by $\lambda_{1}$ and $\lambda_{2}$. Finally, for any state
described by a realistic local theory, the Bell inequality reads
\begin{equation}
\mathcal{B}(\hat{\rho})=2\sqrt{\lambda_{1}+\lambda_{2}}\le2.
\end{equation}

Though the mathematical procedure seems to be complex, $\mathcal{B}(\hat{\rho})$
can be obtained analytically for any given two-qubit state $\hat{\rho}$.
As we will show, $\mathcal{B}(\hat{\rho})$ is just related to two
correlation functions of the system.

\subsection{concurrence}

In order to characterize the properties of the bipartite correlation,
we will further calculate the entanglement concurrence.\citep{def_CC}
Concurrence describes the pairwise entanglement between the two sites.
Firstly, one defines a matrix $\tilde{\rho}$ as $\tilde{\rho}=\hat{\sigma}_{y}\otimes\hat{\sigma}_{y}\hat{\rho}^{*}\hat{\sigma}_{y}\otimes\hat{\sigma}_{y}$,
then the concurrence is $\mathcal{C}=\textrm{max}\{0,\mu_{1}-\mu_{2}-\mu_{3}-\mu_{4}\}$,
where $\mu_{i}$ are the square roots of the eigenvalues of the matrix
$\hat{\rho}\tilde{\rho}$ in decreasing order. One can easily prove
that the concurrence for the state $|\uparrow_{1}\uparrow_{2}\rangle$
is zero, i.e., it is separable, while the concurrence for the state
$|\uparrow_{1}\uparrow_{2}\rangle+|\downarrow_{1}\downarrow_{2}\rangle$
is $1$, i.e., it is maximally entangled.

\subsection{Model}

\begin{figure}
\includegraphics[scale=0.8]{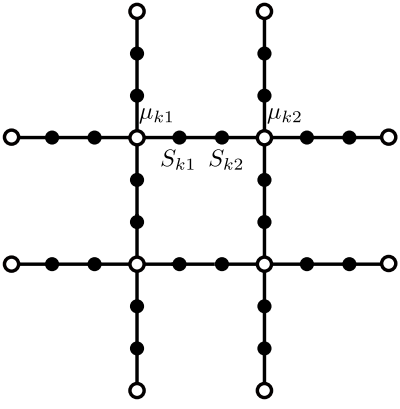}\caption{\label{fig:topo} The topology of the two-dimensional mixed spin model.}
\end{figure}

We consider an exactly soluble two-dimensional mixed spin model firstly
considered by J. Strecka.\citep{2D-mixed_model} The lattice (see
Fig. \ref{fig:topo}) contains two kinds of spins: the Ising spins
$\mu$ and the Heisenberg spins $S$, denoted by white dots and black
dots, respectively. Four continuous spins forms a bond, labeled by
$k$, and different bonds are connected with each other by the Ising
spins. The two adjacent Heisenberg spins on bond $k$ have typical
anisotropic Heisenberg interaction $-J(\Delta S_{k1}^{x}S_{k2}^{x}+\Delta S_{k1}^{y}S_{k2}^{y}+S_{k1}^{z}S_{k2}^{z})$,
with $J$ the coupling parameter and $\Delta$ the anisotropy parameter,
and the adjacent Heisenberg and Ising spins of bond $k$ have an Ising
type interaction, i.e., $-J_{1}S_{k1}^{z}\mu_{k1}^{z}$. Finally,
the totally Hamiltonian of the system is given by

\begin{equation}
\begin{array}{ccl}
H & = & -J\sum_{k}(\Delta S_{k1}^{x}S_{k2}^{x}+\Delta S_{k1}^{y}S_{k2}^{y}+S_{k1}^{z}S_{k2}^{z})\\
 &  & -J_{1}\sum_{k}(S_{k1}^{z}\mu_{k1}^{z}+S_{k2}^{z}\mu_{k2}^{z})
\end{array}
\end{equation}
where the summation is over all bonds.

At zero temperature the system has two phases, i.e., a QAF phase and
an FM phase, and a first-order phase transition separates the two
from each other. The phase boundary is located at $J_{1c}=\frac{\Delta^{2}-1}{2}J$.\citep{2D-mixed_model}
At finite temperatures, the system undergoes a second-order phase
transition from the QAF phase to the paramagnetic (PM) phase for $J_{1}<J_{1c}$.
While for $J_{1}>J_{1c}$, an FM-PM thermal phase transition will be observed. 
In this paper, the parameters are set as $J=1$ and $\Delta=2$,
thus the QPT is located at $J_{1c}=1.5$.

We just pay our attention to the two adjacent Heisenberg spins on
a bond, in which quantum correlation is present. Because of the symmetries
of the system, the reduced density matrix $\hat{\rho}$ of the two-spin
subsystem should be expressed as:\citep{symmetry_rho}

\begin{equation}
\hat{\rho}=\left(\begin{array}{cccc}
u_{+} & 0 & 0 & 0\\
0 & v_{+} & z & 0\\
0 & z & v_{-} & 0\\
0 & 0 & 0 & u_{-}
\end{array}\right),\label{eq:rho}
\end{equation}
where

\begin{equation}
\begin{array}{ccl}
u_{\pm} & = & \frac{1}{4}\pm M_{z}+q_{zz}\\
v_{\pm} & = & \frac{1}{4}\pm\delta S_{z}-q_{zz}\\
z & = & 2q_{xx}
\end{array},
\end{equation}
with $M_{z}=\frac{1}{2}\langle S_{1z}+S_{2z}\rangle$,$\delta S_{z}=\frac{1}{2}\langle S_{1z}-S_{2z}\rangle$,$q_{zz}=\langle S_{1z}S_{2z}\rangle$
and $q_{xx}=\langle S_{1x}S_{2x}\rangle$. According to Horodecki's
formula, the Bell inequality turns out to be

\begin{equation}
\mathcal{B}(\hat{\rho})=\max\{N_{1},N_{2}\}\le2
\end{equation}
with $N_{1}=8\sqrt{q_{zz}^{2}+q_{xx}^{2}}$ and $N_{2}=8\sqrt{2q_{xx}^{2}}$.

Finally, to obtain the Bell function, one just needs to evaluate $q_{zz}$ and $q_{xx}$. 
Based upon the
well-known existing results for two-dimensional Ising model, the solution
of $q_{zz}$ and $q_{xx}$ involves Kambe projection method, the transfer
matrix theory and several powerful spin identities. The calculation
is rather lengthy. Readers who are interested in the method can read
J. Strecka's paper.\citep{2D-mixed_model} Here we just give the
final results as follows

\begin{equation}
\begin{array}{ccc}
q_{xx} & = & \frac{K_{1}+K_{2}}{8}+\frac{K_{1}-K_{2}}{2}\cdot q_{\mu\mu}\\
q_{zz} & = & \frac{K_{3}+K_{4}}{8}+\frac{K_{3}-K_{4}}{2}\cdot q_{\mu\mu}
\end{array},
\end{equation}
where $q_{\mu\mu}$ is just the spin-spin correlation function of
the classical two-dimensional Ising model, which can be found on J. Strecka's paper and the corresponding references.\citep{2D-mixed_model}
The coefficients $K_{i}$ are given by

\begin{equation}
\begin{array}{cll}
K_{1} & = & \frac{\sinh(\beta\frac{1}{2}J\Delta)}{e^{\frac{\beta J}{2}}\cosh\frac{\beta J_{1}}{2}+\cosh(\beta\frac{1}{2}J\Delta)}\\
K_{2} & = & \frac{\sinh(\beta\frac{1}{2}\sqrt{J_{1}^{2}+J^{2}\Delta^{2}})}{e^{\frac{\beta J}{2}}+\cosh(\beta\frac{1}{2}\sqrt{J_{1}^{2}+J^{2}\Delta^{2}})}\cdot\frac{J\Delta}{\sqrt{J_{1}^{2}+J^{2}\Delta^{2}}}\\
K_{3} & = & \frac{\cosh\frac{\beta J_{1}}{2}-e^{-\frac{\beta J}{2}}\cosh(\beta\frac{1}{2}J\Delta)}{\cosh\frac{\beta J_{1}}{2}+e^{-\frac{\beta J}{2}}\cosh(\beta\frac{1}{2}J\Delta)}\\
K_{4} & = & \frac{e^{\frac{\beta J}{2}}-\cosh(\beta\frac{1}{2}\sqrt{J_{1}^{2}+J^{2}\Delta^{2}})}{e^{\frac{\beta J}{2}}+\cosh(\beta\frac{1}{2}\sqrt{J_{1}^{2}+J^{2}\Delta^{2}})}
\end{array}.
\end{equation}

\section{Bell function and phase transitions}

In this section, we discuss the ability of Bell function in detecting
the phase transitions in the model.

\begin{figure}
\includegraphics[scale=0.7]{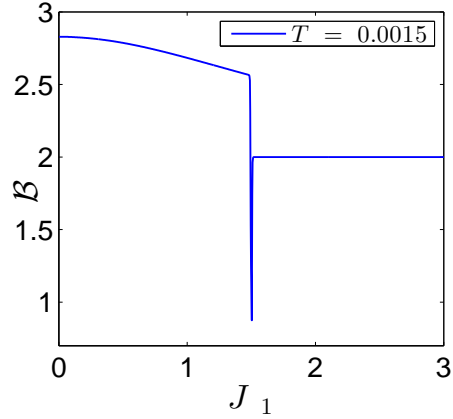}\caption{\label{fig:QPT} The Bell function $\mathcal{B}$ at $T=0.0015$.
It shows a sudden-change at $J_{1}=1.5$.}
\end{figure}

Firstly, we consider the QPT of the system. At zero temperature (and
low temperatures), the system is in the QAF phase for $J_{1}<1.5$,
and in the FM phase for $J_{1}>1.5$. In Fig. \ref{fig:QPT} we show
the Bell function as a function of $J_{1}$ at $T=0.0015$. The Bell
function shows a sudden-change at $J_{1}=1.5$. The discontinuity
of $\mathcal{B}$ should result from the sudden-change of the density
matrix, thus can be regarded as a reliable signal for a first-order
phase transition. In addition, in the QAF phase, the value of the
Bell function is larger than 2, thus the Bell inequality is violated.
While in the FM phase, the Bell inequality is never violated.

\begin{figure}
\includegraphics[scale=0.6]{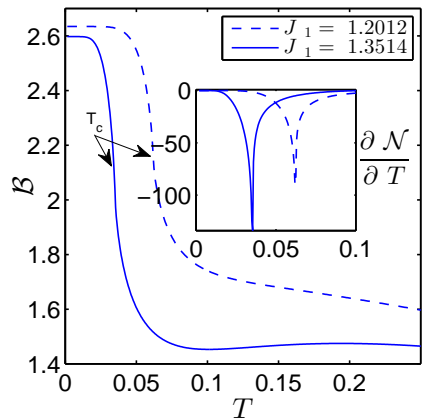}\caption{\label{fig:QAF-PM} The behavior of Bell function in the QAF-PM thermal
phase transition for several $J_{1}$. The Inset is the derivative
of the Bell function with respect to the temperature. The derivative
of Bell function is divergent at the critical temperature $T_{c}=0.035$
for $J_{1}=1.351$ and $T_{c}=0.063$ for $J_{1}=1.201$.}
\end{figure}

Secondly, we study the behavior of the Bell function in the thermal
phase transitions at finite temperature. The system has a QAF-PM transition
for $J_{1}<1.5$, and an FM-PM transition for $J_{1}>1.5$. In Fig.
\ref{fig:QAF-PM} we show the temperature dependence of the Bell function
in the QAF-PM transition. The derivative of $\mathcal{B}$ is divergent
at some critical temperatures. Divergence of $\frac{\partial\mathcal{B}}{\partial T}$
should result from continuous and dramatic change of the density matrix,
thus can be regarded as a reliable signal for second-order phase transitions.
In fact, we have checked that the location of the divergence of the
Bell function is indeed perfectly consistent with the divergent point
of specific heat, thus the Bell function indeed identifies exactly
the location of the thermal phase transition. In addition, in low-temperature
region, the Bell function is larger than 2. Then, as the increase
of the temperature, the Bell function decreases gradually. And finally,
the Bell inequality is not violated at high-temperature regions. It
shows that the nonlocality present at low temperatures is finally
destroyed by thermal fluctuation.

\begin{figure}
\includegraphics[scale=0.6]{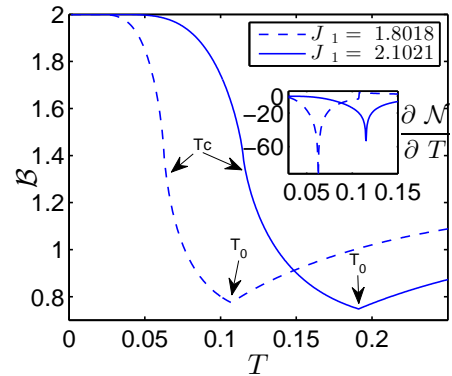}\caption{\label{fig:FM-PM} The behavior of Bell function in the FM-PM thermal
phase transition for several $J_{1}$. The Inset is the derivative
of the Bell function with respect to the temperature. The derivative
of Bell function is divergent at the critical temperature $T_{c}=0.063$
for $J_{1}=1.802$ and $T_{c}=0.115$ for $J_{1}=2.102$.}
\end{figure}

In Fig. \ref{fig:FM-PM} we plot the behavior of Bell function in
the FM-PM thermal phase transition. Just as in the QAF-PM transition
in Fig. \ref{fig:QAF-PM}, the derivative of $\mathcal{B}$ shows
a divergence at the second-order phase transition point. In addition, the
derivative of Bell function shows a discontinuity at $T_{0}=0.107$
for $J_{1}=1.802$ and $T_{0}=0.191$ for $J_{1}=2.102$. Detailed
study manifests that it just results from the max function in calculating
the Bell function, rather than a phase transition.

In fact, according to Sec. 2, the Bell function is related to the
spin-spin correlation functions as
\begin{equation}
\mathcal{N}=\max\{8\sqrt{q_{zz}^{2}+q_{xx}^{2}},8\sqrt{2q_{xx}^{2}}\}.\label{eq:nonlocality_2D}
\end{equation}
At the phase transition points, $q_{zz}$ or $q_{xx}$ can be singular.
This is the mathematical reason why $\mathcal{N}$ can be used to
detect the phase transitions in this model. It needs mention that
the density matrix in Eq. (\ref{eq:rho}) contains four free parameters
$M_{z}$, $\delta S_{z}$, $q_{zz}$ and $q_{xx}$. The singularity
of any of these four parameters would indicate a phase transition
of the system. In fact, $M_{z}$ and $\delta S_{z}$ are just the
traditional order parameters for the FM-PM and QAF-PM phase transitions,
respectively. However, $\mathcal{N}$ just contains the information
of $q_{zz}$ and $q_{xx}$. It needs mention that the concurrence is
a function of $q_{zz}$, $q_{xx}$, and $M_{z}$, i.e., $\mathcal{C}=2\max\{0,2\vert q_{xx}\vert-\sqrt{(\frac{1}{4}+q_{zz})^{2}-M_{z}^{2}}\}$.\citep{SunEpl}
From a mathematical point of view, one may conclude that concurrence
is better than the nonlocality in detecting phase transitions, since
it contains more parameters in $\hat{\rho}$. However, it
turns out that the concurrence $\mathcal{C}$ is zero in the FM-PM
transition,\citep{SunEpl} while $\mathcal{N}$ is just zero in a
very narrow parameter space and it indeed offers a singularity at
the critical temperature. Concurrence can only detect phase transitions
in entangled states, while the Bell function can capture signals of
phase transition in non-local states (Fig. \ref{fig:QAF-PM}) and
local states (Fig. \ref{fig:FM-PM}).

Furthermore, the max function in Eq. (\ref{eq:nonlocality_2D}) induces
a non-physical singularity, i.e., the discontinuity of $\frac{\partial\mathcal{N}}{\partial T}$
in Fig. \ref{fig:QAF-PM}. This kind of singularity should not be
regarded as a reliable signal of phase transitions, since it can results
from either the singularity of the density matrix, or merely the max
function. Thus, if $\frac{\partial\mathcal{N}}{\partial T}$ turns
out to be discontinuous in some models, we need to be very careful
to judge whether it is a critical temperature or not. This kind of
discontinuity has been observed in other measures of correlation such
as concurrence and discord.\citep{re_examin_CC-1,SunEpl}

\section{Properties of bipartite correlations}

\begin{figure}
\includegraphics[scale=0.45]{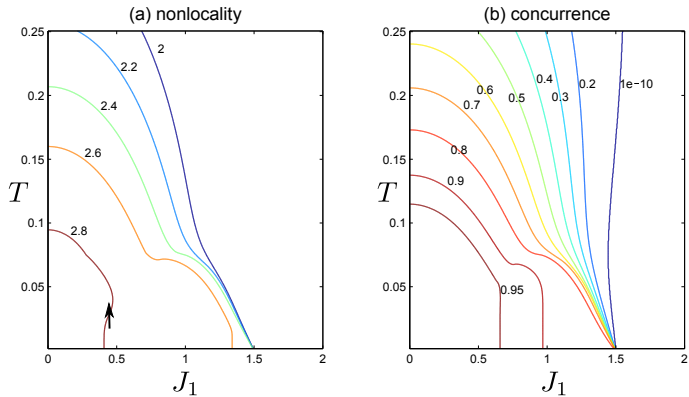}\caption{\label{fig:contour} Contour maps for the (a) nonlocality and (b)
concurrence in the $T-J_{1}$ plane.}
\end{figure}

We calculate the Bell function and the concurrence in the $T-J_{1}$
plane, and the corresponding contour maps are shown in Fig. \ref{fig:contour}.

The figure shows that weaker coupling and lower temperature will result
in higher quantum correlation in this model. One can see that high
nonlocality or high entanglement is present in the bottom left corner
of the maps. The effect of coupling constant on quantum correlation
is model-dependent. Nevertheless, the relationship between temperature
and quantum correlation is more general, because thermal fluctuation
will destroy quantum correlation. From Fig. \ref{fig:contour} one
sees that for most $J_{1}$, as the temperature increases, both the
nonlocality and concurrence will decrease gradually.

However, we find the nonlocality can be enlarged by thermal fluctuation
in some parameter space. As indicated by the black arrow in Fig. \ref{fig:contour},
the contour line for $\mathcal{N}=2.8$ bends to the right. As a result,
for an appropriate fixed $J_{1}$, for example $J_{1}=0.48$, when
the temperature increases from zero, the Bell function will cross
the $\mathcal{N}=2.8$ contour line from below, which means that the
nonlocality is enhanced by thermal fluctuation. In the high-entanglement
region in Fig. \ref{fig:contour} (b), e.g., in the vicinity of the
$\mathcal{C}=0.95$ contour line, the concurrence is not enhanced
by thermal fluctuation. The different behavior between nonlocality
and entanglement shows that the two concepts indeed capture different
aspects of quantum correlation.

\begin{figure}
\includegraphics[scale=0.45]{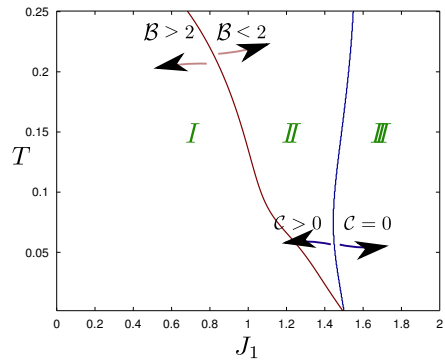}\caption{\label{fig:three-regions} (Color Online) The $T-J_{1}$ plane is
divided into three regions by analyzing the presence/absence of nonlocality
and entanglement. }
\end{figure}

In Fig. \ref{fig:three-regions}, we divide the $T-J_{1}$ plane into
several regions by considering the form of bipartite correlation in
the system. The red line is just the contour line for $\mathcal{N}=2$,
on the left side of which the system is non-local. In addition, the
blue line is the contour line of $\mathcal{C}=0$, on the left side
of which the system is entangled. Finally, the whole $T-J_{1}$ plane
is divided into three regions, labeled as $\mathrm{I}$, $\mathrm{II}$,
$\mathrm{III}$. In region $\mathrm{I}$, both entanglement and nonlocality
are present in the system. In region $\mathrm{II}$, the correlation
is in the form of entanglement, rather than nonlocality. In region$\mathrm{III}$,
where the FM-PM phase transition happens, there is no nonlocality
or entanglement in the system. However, in a previous study,\citep{SunEpl}
we have found that quantum correlation is indeed present in the FM-PM
phase transition by studying quantum discord. This result shows that
the quantum correlation in region $\mathrm{III}$ is very novel, which
cannot be characterized by entanglement or nonlocality. Quite recently,
some other measures of bipartite correlation have been developed,
for example, the so-called measure $Q$ for $2\times d$ systems.\citep{measure-Q}
These measures may offer further information about the quantum correlation
in region $\mathrm{III}$.

\section{Summaries and discussions}

In this paper, we have studied the bipartite correlation in a two-dimensional
mixed spin system. We have used the Bell inequality to investigate the nonlocality
in the system.

Firstly, Bell function captures the signals for both the quantum phase
transition and thermal phase transitions of the model. In the vicinity
of zero temperature, the Bell function $\mathcal{N}$ shows a sudden-change
at the transition point $J_{1}=1.5$. In addition, at finite temperatures,
$\frac{\partial\mathcal{N}}{\partial T}$ is divergent at the transition
temperatures. The Bell function captures not just the location, but
also the order, of these phase transitions.

Secondly, we find a novel type of quantum correlation, which cannot
be characterized by entanglement or nonlocality. In region $\mathrm{III}$
of Fig. \ref{fig:three-regions}, the Bell inequality is not violated,
and the concurrence is zero. However, a previous study shows that
there is indeed quantum correlation present in the system.\citep{SunEpl}
Recently developed tools such as the measure $Q$ may be helpful to
describe further physical picture for this kind of quantum correlation.\citep{measure-Q}

Thirdly, we find the nonlocality can be enlarged by thermal fluctuation
in the anti-ferromagnetic phase(see Fig. \ref{fig:QAF-PM}). As in-depth discussion is needed,
let's review several related papers first. In a one-dimensional
diamond-like quantum spin model, the concurrence increases from $0$
to finite values when thermal fluctuation is increased gradually.\citep{finite-temperature-entanglement}
Similarly, for just the two-dimensional model studied in this paper,
the discord, which is $0$ at zero temperature, is also enlarged by
thermal fluctuation.\citep{SunEpl} It needs mention that, in these
two situations, the ground states are ferromagnetic, i.e. $|\uparrow\uparrow\uparrow...\rangle$,
and all spins are parallel to each other, thus the quantum correlation
is exactly zero at zero temperature. Then, as the increase of the
temperature, excited states are mixed into the density matrix. These
excited states are usually complicated, and it is expected for them
to bring some quantum correlation into the system. That's why thermal
fluctuation can enhance correlation in ferromagnetic phases. However,
our finding in this paper cannot be explained by the above picture.
The ground state of the system is anti-ferromagnetic in the parameter space $J_1 < J_{1c}$. Consequently,
the system is in a highly non-local state. Thus the enhancement of
nonlocality by thermal fluctuation in Fig. \ref{fig:QAF-PM} is
unusual. A possible approach to understand the behavior is to calculate
low-lying states $|e_{i}\rangle$ explicitly, and analyze the mixing
process $\sum_{i}e^{-\beta e_{i}}|e_{i}\rangle\langle e_{i}|\rightarrow\hat{\rho}$
as the change of the temperature. We'd like to mention that any qualitative
discussion will be untenable, since the entanglement concurrence is
not enlarged by the thermal fluctuation in the same parameter region.
It suggests that quantum correlation phenomena, despite being extensively
studied, is worthy of further investigation.

\section*{Acknowledgments}

The research was supported by the National Natural Science Foundation
of China (No. 11204223). This work was also supported by the Talent
Scientific Research Foundation of Wuhan Polytechnic University (Nos.
2012RZ09 and 2011RZ15).

\section*{References}

\appendix

\end{document}